\documentclass[%
 amsmath,amssymb,
 aps,
]{revtex4-2}

\def \bF {\pmb{F}}
\def \bH {\pmb{H}}

\def \bx {\pmb{x}}

\def \bz {\pmb{z}}

\usepackage{graphicx}
\usepackage{dcolumn}
\usepackage{bm}
\usepackage[dvipsnames]{xcolor}
\usepackage[normalem]{ulem}
\usepackage[title]{appendix}
\usepackage{comment}
\usepackage{float}
\usepackage{mathtools}

\usepackage{hyperref}

\graphicspath{{./figures/}}
\begin{document}
		
			\title{\bf {Failure of the simultaneous block diagonalization technique applied to complete and cluster synchronization of random networks}}

	\begin{abstract}
We discuss here the application of the simultaneous block diagonalization (SBD) of matrices to the study of the stability of both complete and cluster synchronization in random (generic) 
networks. For both problems, we {define indices that measure success (or failure) of application of the SBD technique in decoupling the stability problem into problems of lower dimensionality. We then see that in the case of random networks}
 the extent of the dimensionality reduction achievable is the same as that produced by application of a trivial transformation. \\

	\end{abstract}

			\author{ Shirin Panahi, Nelson Amaya, Isaac Klickstein, Galen Novello, Francesco Sorrentino \\
			University of New Mexico, Albuquerque, US 80131}

	\maketitle


\section{Introduction} \label{s:intro}

The mathematics literature has dealt with the fundamental problem of simultaneous block diagonalization (SBD) of a set of matrices  \cite{UHLIG1973281,Maehara2010-1,Maehara2010-2,Murota2010,Maehara2011}. The first paper where this technique was applied to network synchronization is \cite{Ir:So}, which focused on complete synchronization of networks with nodes connected through two or more coupling functions. More recently this technique has been applied to cluster synchronization of networks \cite{Zhang2020}. 
The use of this technique reduces the stability problem to a number of subproblems of smallest dimension. 
The original Master Stability Function (MSF) derivation of \cite{Pecora1998}  decouples the stability problem for any $N$-dimensional matrix corresponding to an undirected network into $N$ independent blocks, where each block coincides with one of the matrix eigenvalues. 
However, it is unclear what the extent of the dimensionality reduction obtained from application of the SBD technique can be.

Here we characterize the extent of this dimensionality reduction when the SBD approach is applied to generic networks, where by a generic network we mean a `typical' network that is produced by a random process such as the Erd{\H{o}}s-R{\'e}nyi network generation algorithm \cite{erdHos1960} or the configuration model \cite{Mo:Re95}. 
{Random networks are broadly studied in the literature as fundamental and paradigmatic models for the structure and dynamics of complex systems \cite{Report}. Previous work has investigated random networks in the context of epidemics \cite{pastor2001epidemic,marder2007dynamics,pastor2015epidemic}, percolation \cite{achlioptas2009explosive,friedman2009construction}, resilience to attacks and failures \cite{guillaume2004comparison,liu2012cascading}, games \cite{devlin2009evolution}, network synchronization \cite{restrepo2006synchronization} and control \cite{liu2011controllability}.}
{It is therefore important to characterize both complete and cluster synchronization for this class of networks.} We show that application of the SBD reduction to these random networks does not lead to a beneficial reduction of the stability problem, either in the case of complete synchronization or cluster synchronization. Nonetheless, we do not mean that the technique is not useful. However, it points out that its usefulness is limited to the non-generic case, for which the reduction can sometimes be very significant\cite{Ir:So,Zhang2020}.

{Our paper is structured as follows: 
In sections \ref{II} and \ref{III} we provide the mathematical background for the method we use to compute the SBD. Our main results are presented in Sections \ref{Ap1} and \ref{Ap2}, which discuss the cases of complete and cluster synchronization respectively.  In those sections, we define indices to measure the extent of the dimensionality reduction resulting from the application of the SBD algorithm. In the case of randomly constructed networks, we see that the index value often equals zero to demonstrate certain limitations of the method. In section \ref{VI}, we present a discussion on the relevance of our findings in applying the SBD to randomly constructed networks. Lastly, the conclusions are given in section \ref{s:conclusion}. }

\section{Simultaneous Block Diagonalization of Matrices}\label{II}

The problem of simultaneous block diagonalization can be formalized as follows: given a set of $N \times N$ matrices $A^{(1)}, ... , A^{(M)}$ find an $N \times N$ orthogonal matrix $P$ such that the matrices $P^{T}A^{(k)}P$ have a common block-diagonal structure for $k = 1, ... , M$. It should be noted that such a block-diagonal structure is not unique in at least two senses:  first, the blocks may be permuted, resulting in block diagonal decompositions that are isomorphic; second, the matrices corresponding to certain blocks may be further refined into smaller blocks, resulting in a structure that is fundamentally different. A block diagonal structure with smaller blocks is considered to be finer and the finest SBD (FSBD) is beneficial in that it provides the simplest elements in the decoupling of systems as described above. 

There are two different but closely related theoretical frameworks with which we
can address our problem of finding a block-diagonal decomposition for a finite set
of given $N \times N$ real matrices. The first is group representation theory \cite{Miller1973,Serre1977} which relies on group symmetries and ensures a degree of universality in a SBD. The second is matrix $*$-algebras \cite{Wedderburn1934} which are not only necessary to answer the fundamental
theoretical question of the existence of such a finest block-diagonal decomposition but
also useful in its computation. Indeed, existence can be justified through the structure theorem of $*$-algebras \cite[Theorem 5.4]{Kojima1997} and this structure has also been utilized to formulate algorithms for computing the SBD of $A^{(1)}, ... , A^{(M)}$. In particular, our approach appeals to this structure, but it should be noted that both frameworks have been utilized in the literature \cite{Bai2009,Klerk2010,Gatermann2004,Riener2013,Kanno2001}.

In what follows we write
\begin{equation}
    P=\mathcal{SBD}(A^{(1)},A^{(2)},...,A^{(M)})
\end{equation}
to indicate that the transformation yields
\begin{equation}
    P^T A^{(k)} P=B^{(k)}, \quad k=1,...,M,
\end{equation}
where all the matrices $B^{(k)}$, $k=1,...,M$ share the same finest block diagonal form,
\begin{equation}
    B^{(k)}=\bigoplus_j B_j^{(k)},
\end{equation}
with the blocks $B_j^{(k)}$ all having the same sizes for $k=1,...,M$ and not being further reducible {by a simultaneous transformation.}

\section{Procedure to Determine $P$}\label{III}

Here we describe the procedure to compute the FSBD for a set of {$M$ symmetric matrices denoted $A^{(k)}$, $k = 1,2,\ldots,M$, previously published \cite{Maehara2011}}.
First, we find a matrix $U$ that simultaneously commutes with each matrix $A^{(k)}$ \cite{Maehara2011}, that is, $[A^{(k)},U] = A^{(k)} U - U A^{(k)} = O_n$, $k = 1,\ldots,M$, where $O_n$ is the $n$-by-$n$ matrix of all zeroes.

{Define the vectorizing function ${\mathcal{V}} : \mathbb{R}^{n \times m} \mapsto \mathbb{R}^{nm}$ to take as input an $n$-by-$m$ matrix and return a vector by stacking each of the matrix's columns on top of each into a vector of length $nm$.}
{For two matrices $A \in \mathbb{R}^{n \times m}$ and $B \in \mathbb{R}^{p \times \ell}$, let $A \otimes B \in \mathbb{R}^{np \times m\ell}$ denote the Kronecker product.}
As the commutator equation is linear in $U$, it can alternatively be expressed as a matrix-vector product.
{
\begin{equation}\label{eq:vec_commutator}
\begin{aligned}
    {\mathcal{V}}(A^{(k)} U - U A^{(k)}) &={\mathcal{V}}(A^{(k)}U{)} - {\mathcal{V}}(U A^{(k)})\\
    &={\mathcal{V}}(O_n), \quad k = 1,2,\ldots,M
\end{aligned}
\end{equation}
}
{The vectorizing function applied to a matrix product can be expressed as a matrix-vector product, where for $A \in \mathbb{R}^{n \times m}$ and $B \in \mathbb{R}^{m \times p}$, the product ${\mathcal{V}}(AB) = (I_p \otimes A) {\mathcal{V}}(B) = (B^T \otimes I_n) {\mathcal{V}}(A)$ (see Proposition 7.1.9 in \cite{Bernstein2009}).}
{Apply these identities to Eq. \eqref{eq:vec_commutator}, and define} $P^{(k)} {\mathcal{V}}(U) = (I_n \otimes A^{(k)} - (A^{(k)})^{T} \otimes I_n) {\mathcal{V}}(U) = \boldsymbol{0}_{N^2} $ where $\boldsymbol{0}_{N^2}$ is the vector of all zeros of length $N^2$.
{To find $U$, we look for a vector in the intersection of the nullspaces of $P^{(k)}$ for $k = 1,2,\ldots,M$, that is, a vector ${\mathcal{V}}(U) \in \bigcap_{k=1}^M \mathcal{N}(P^{(k)})$.}
{This can be accomplished in two steps by first noting that for a matrix $A \in \mathbb{R}^{n \times m}$ $\mathcal{N}(A) = \mathcal{N}(A^TA)$ (see Theorem 2.4.3 in \cite{Bernstein2009}) and second, for a set of $M$ positive semi-definite matrices $B^{(j)}$, $j = 1,2,\ldots,M$, $\mathcal{N}\left(\sum_{j=1}^M B^{(j)} \right) = \bigcap_{j=1}^M \mathcal{N}(B^{(j)})$ (see Fact 8.7.3 in \cite{Bernstein2009}). }
As the matrix $P^{(k)}$ may not be positive semi-definite, {the vectorized commutator operation} is pre-multiplied by $(P^{(k)})^T$, so that the matrix $(P^{(k)})^T P^{(k)}$ is symmetric and positive semi-definite.
Create the matrix $S = \sum_{k=1}^M (P^{(k)})^T P^{(k)}$ so that if a vector ${\mathcal{V}}(U)$ is in the nullspace of $S$, {it lies in the intersection of the nullspaces of $P^{(k)}$, and thus} it also commutes with all $A^{(k)}$, $k = 1,\ldots,M$.\\
%
\indent 
To determine the nullspace of the matrix $S \in \mathbb{R}^{N^2 \times N^2}$, which by construction is positive semi-definite, we find the eigenvectors of $S$ corresponding to eigenvalues equal to zero.
While $S$ is large and dense, its special structure makes finding a few extremal eigenvalues and eigenvectors feasible even when $N$ is large {by using the Lanczos method \cite{Saad2013} which only requires a function to compute matrix-vector products.}
Note that while a matrix-vector product, $S \boldsymbol{u}$, requires $N^4$ operations (remember $S \in \mathbb{R}^{N^2 \times N^2}$), it can equivalently be computed using nested commutation operations requiring $4N^3$ operations. {To see this reduction, we can break the matrix-vector product into individual contributions from each $P^{(k)}$.
\begin{equation}\label{eq:nested_commutators}
\begin{array}{cc}
    S \boldsymbol{u} = \sum_{k=1}^M P^{(k)^T} P^{(k)} \boldsymbol{u} = \sum_{k=1}^M P^{(k)^T} {\mathcal{V}}\left([A^{(k)},\text{mat}(U)]\right) =\\ \sum_{k=1}^M {\mathcal{V}}\left(\left[A^{(k)^T},\left[ A^{(k)},U \right] \right] \right) 
\end{array}
\end{equation}
Evaluating a commutator requires two $N$-by-$N$ matrix products which each requires $2N^3$ operations.
In total, for $M$ twice nested commutators, the total work required is $4MN^3$ operations, which for $M \ll N$, is a significant reduction as compared to constructing $S$ explicitly.}
This can be demonstrated with the following steps to compute ${\mathcal{V}}(Y) = S{\mathcal{V}}(U)$.
\begin{enumerate}
    \item Initialize $Y = O_N$ to be the $N$-by-$N$ matrix of zeroes
    \item For $k = 1,2,\ldots,M$
    \begin{enumerate}
        \item $\hat{U} = A^{(k)} U - U A^{(k)}$
        \item $Y \gets Y + A^{(k)^T} \hat{U} - \hat{U} A^{(k)^T}$
    \end{enumerate}
    \item Return ${\mathcal{V}}(Y)$
\end{enumerate}
The computational complexity can further be reduced if each of the matrices $A^{(k)}$ is sparse with average density $\rho \in [0,1]$ {(defined as the number of nonzero entries divided by $N^2$)} so computing the matrix-matrix products requires in average only $4 \rho N^3$ operations.\\
\indent
The dominant computational complexity of each step of the Lanczos algorithm is the matrix vector product \cite{Saad2013} which we have shown can be computed in $4 \rho M N^3$ operations, far more cheaply than the $N^4$ operations if $S$ did not have its special structure.
Due to the iterative nature of the Lanczos algorithm, it is unknown \textit{a priori} the number of iterations required to compute the eigenvalue/eigenvector pairs.
Nonetheless, unless the number of iterations required is on the order of $N^2$ or larger, the Lanczos algorithm is more efficient than constructing $S$ explicitly and finding the eigenvalue/eigenvectors pairs using a standard tridiagonalization approach for dense symmetric eigenvalue problems.\\
\indent 
Let $\boldsymbol{v}_k \in \mathbb{R}^{N^2}$, $k = 1,2,\ldots,n_{ev}$, be the $n_{ev}$ eigenvectors found corresponding to the eigenvalues equal to zero of $S$, {each of which lies in the intersection of the nullspaces of $P^{(k)}$, $k = 1,2,\ldots,M$}. {To select a random vector in the intersection of the nullspaces, create the vector ${\mathcal{V}}(U) = \sum_{k=1}^{n_{ev}} c_k \boldsymbol{v}_k$ where we uniformly at random select } $c_k \in [-1,1]$, $k = 1,2,\ldots,n_{ev}$, and scale them such that $\sum_{k} c_k^2 = 1$.
The resulting matrix $U$ satisfies all of the commutation relations, $[A^{(k)},U] = O_N$, as does $U^T$ because each $A^{(k)}$ is symmetric.
With this fact, the symmetric matrix $\frac{1}{2} (U+U^T)$ also commutes, $\left [A^{(k)}, \frac{1}{2}(U+U^T) \right]$, $k = 1,2,\ldots,M$.
Finally, to find the matrix $P$ that simultaneously block diagonalizes each of the $A^{(k)}$, $k=1,2,\ldots,M$, compute the eigenvectors of $\frac{1}{2}(U+U^T)$, and store them as the columns of $P$.
The proof of the correctness is extensive and beyond the scope of this paper but can be found in \cite{Maehara2011}.\\
\indent 
A related problem \cite{Maehara2011} is to find a transformation $P$ that does not exactly simultaneously block diagonalize all of the matrices $A^{(k)}$, $k = 1,2,\ldots,M$, but rather results in matrices with off-diagonal blocks with entries with magnitude of the order $\epsilon > 0$.
The process {is validated by Lemma 4.1 in \cite{Maehara2011} } and proceeds exactly as before except now rather than finding the eigenvectors associated with eigenvalues equal to zero, instead, the Lanczos method is used to find eigenvalues of $S$ less than $\epsilon$ along with their eigenvectors.
After this, with the eigenvectors $\boldsymbol{v}_k$, $k = 1,2,\ldots,n_{ev}$, the same steps are taken to compute $U$, extract its symmetric part $\frac{1}{2}(U+U^T)$, and find the eigenvectors of the result.

We make available our code to compute the FSBD of a set of symmetric matrices -- see \cite{Code}.

\section{Application of the SBD technique to complete synchronization of networks with different types of connections } \label{Ap1}

The time evolution of a network of dynamical systems coupled through different types of connections is described by the following set of equations:
\begin{equation}
\dot{\bx}_{i}(t) = \bF(\bx_{i}(t))+\sum_{k=1}^{M}\sum_{j=1}^{N}A_{ij}^{(k)}[\bH^{(k)}(\bx_{j}(t))-\bH^{(k)}(\bx_{i}(t))] \quad  i=1, \cdots, N
\label{Eq1}
\end{equation}
where $\bx_i(t)$ and $\bF(\bx_i(t))$ represent the $m$-dimensional state vector and dynamical function of the system located at node $i$, respectively. The network nodes are coupled through different coupling functions $\bH^{(k)}$, $k=1, \cdots M$. 
The network connectivity associated to each coupling function is described by the adjacency matrix $A^{(k)}$, where $A_{ij}^{(k)}=A_{ji}^{(k)}>0$ if there is a connection between nodes $i$ and $j$ and $A_{ij}^{(k)}=A_{ji}^{(k)}=0$ otherwise. The above set of equations can be 
rewritten,
\begin{equation}
\dot{\bx}_{i}(t) = \bF(\bx_{i}(t))+\sum_{k=1}^{M}\sum_{j=1}^{N}L_{ij}^{(k)}\bH^{(k)}(\bx_{j}(t)) \quad  i=1, \cdots, N,
\label{Eq2}
\end{equation}
where the Laplacian matrices $L^{(k)}$ have entries  $L_{ij}^{(k)}=A_{ij}^{(k)}>0$ for $j \neq i$ and  $L_{ii}^{(k)}=-\sum_{j \neq i} L_{ij}^{(k)}$, $k=1,...,M$. Note that all of the rows of the Laplacian matrices $L^{(k)}$ sum to zero, $k=1,...,M$. The synchronization manifold $\bx_{1}=\bx_{2}= \cdots = \bx_{N}$ is an invariant subspace for the set of Eqs.\ \eqref{Eq2}. The dynamics on this manifold, which corresponds to \textit{complete synchronization}, 
$
\bx_{1}(t)=\bx_{2}(t)= \cdots = \bx_{N}(t)= \bx_{s}(t)$
obeys the equation of an uncoupled system, 
\begin{equation}
\bx_{s}(t)=\bF(\bx_{s}(t)).
\end{equation}

To investigate the stability of the complete synchronous state, we study the dynamics of a small perturbation from the synchronous solution $(\bx_{i}(t)=\bx_{s}(t)+\delta \bx_{i}(t))$. The synchronous state is stable if the perturbations approach $0$ for large $t$. The linearized system of equations can be written,
\begin{equation}
  \begin{aligned}
    \delta \dot{ \bx}_{i}(t) &= D\bF(\bx_{s}(t))\delta \bx_{i}(t)\\
    &+ \sum_{k=1}^{M}\sum_{j=1}^{N}L_{ij}^{(k)}D\bH^{(k)}(\bx_{s}(t))\delta \bx_{j}(t) \quad i=1, \cdots, N
  \label{Eq4}
  \end{aligned}
\end{equation}
By stacking together all the perturbations in an   $mN$-dimensional vector $\bz=[\delta \bx_{1}^T,\delta \bx_{2}^T,\cdots,\delta \bx_N^T]^T$, the set of  Eqs.\ \eqref{Eq4} can be rewritten in vectorial form,
\begin{equation}
\dot{\bz}(t) = \left[ I_{N} \otimes D\bF(\bx_{s}(t)) +\sum_{k=1}^{M}L^{(k)} \otimes D\bH^{(k)}(\bx_{s}(t)) \right] \bz(t)
\label{Eq5}
\end{equation}

One observation is that by construction the set of Laplacian matrices $L^{(1)},L^{(2)},...,L^{(M)}$ all share one common eigenvector $[1,1,...,1]/\sqrt{N}$, with associated eigenvalue $0$. It follows that we can define an orthogonal transformation $\tilde{P}$ leading to a trivial simultaneous block diagonalization (TSBD)
\begin{equation}
    \tilde{P}^T L^{(k)} \tilde{P}= 0 \oplus B_{\mbox{orth}}, \quad k=1,...,M, 
\end{equation}
where the block $B_{\mbox{orth}}$ is $(N-1)$-dimensional.
Hence, there will be a large block produced by the TSBD with dimension $N-1$. For example, $\tilde{P}$ could be taken as the matrix whose columns are the eigenvectors of any of the matrices $L^{(1)}$, $L^{(2)},...,L^{(M)}$ \cite{HYP}.
One would hope that calculation of the FSBD  for the set of matrices $L^{(1)},L^{(2)},...,L^{(M)}$ leads to a finer block-diagonalization than the TSBD.


Here, for simplicity and without loss of generality, we focus on the case of $M=2$ different connection types, for which Eq.\ \eqref{Eq5} becomes,
\begin{equation}
\begin{aligned}
\dot{\bz}(t) &= \left[ I_{N} \otimes D\bF(\bx_{s}(t)) + L^{(1)} \otimes D\bH^{(1)}(\bx_{s}(t)) \right. \\
&+ \left. L^{(2)} \otimes D\bH^{(2)}(\bx_{s}(t)) \right]  \bz(t)
\label{Eq6}
\end{aligned}
\end{equation}
We attempt to break the stability of problem \eqref{Eq6} into a set of independent lower-dimensional equations. To this end, we seek for a transformation that leads to decoupling the set of Eqs.\ \eqref{Eq6}, by simultaneously block diagonalizing $L^{(1)}$ and $L^{(2)}$.  Special instances of this problem have been studied in Ref.\ \cite{HYP} which obtained three different conditions under which the problem with $mN$-dimension can be broken into a set of $(N-1)$ problems of dimension $m$ each. Moreover, Ref. \cite{Ir:So} has introduced the general framework in which the SBD technique is applied to network synchronization.

We now compute $P=\mathcal{SBD}(L^{(1)},L^{(2)})$ and rewrite
 Eq.\ \eqref{Eq6} as follows:
\begin{equation}
  \begin{aligned}
    \dot{\pmb{\eta}}(t) &=  \left[I_{N} \otimes D\bF(\bx_{s}(t)) + (P^T L^{(1)}P) \otimes D\bH^{(1)}(\bx_{s}(t))\right.\\
    &+\left. (P^T L^{(2)}P) \otimes D\bH^{(2)}(\bx_{s}(t)) \right] \pmb{\eta}(t),
\label{Eq7}
\end{aligned}
\end{equation}
where $\pmb{\eta}(t) = (P^T \otimes I_{m})  \bz(t)$.

As stated before, $P^T L^{(i)}P=\oplus _{j=1}^{n} B_j^{(i)}$, where all the matrices $B_j$ have the same block-diagonal form. Therefore, Eqs.\ \eqref{Eq7} can be decoupled as follows,
\begin{equation}
  \begin{aligned}
    \dot{{\pmb \eta}_{i}}(t) &= \left[ I_{D_{i}} \otimes D\bF(\bx_{s}(t))  + B_{i}^{(1)} \otimes D\bH^{(1)}(\bx_{s}(t)) \right. \\
    &+ \left. B_{i}^{(2)} \otimes D\bH^{(2)}(\bx_{s}(t)) \right] {\pmb \eta}_{i}(t),
  \end{aligned}
\label{Eq8}
\end{equation}
where $D_{i}$ is the block-dimension of $B_i$, $\sum_i D_i=N$. 

We note that for a given value of $i$ ($i=1$) we obtain scalar blocks $B^{(1)}_1=B^{(2)}_1=0$, which are associated with a perturbation parallel to the synchronization manifold. Therefore, to analyze the stability of the synchronous solution, we only need to assess Eq. \eqref{Eq8} for the remaining $i>1$ transverse blocks.
 
 \subsection{Performance of the SBD technique applied to complete synchronization}
 
To examine the extent of the reduction provided by the SBD method, the following index is introduced
\begin{equation}
0 \leq I_{d}=1-\frac{Z-1}{N-2} \leq 1
\label{Index}
\end{equation}
where $Z$ is the maximum block dimension in both $P^T L^{1} P$ and $P^T L^{2} P$. The best possible performance of the SBD is achieved for $I_d=1$, corresponding to all the blocks having dimension $Z=1$. On the other hand, if the maximum block dimension $Z=N-1$, the index $I_d=0$, which corresponds to the same reduction achievable with the TSBD. {We define application of the SBD technique to be a success (a failure) for large (low) values of $0 \leq I_d \leq 1$.}

We examine the performance of the SBD method in reducing {the dimension of} the problem of complete synchronization for three different network {classes}: (i) Erd{\H{o}}s-R{\'e}nyi (ER) random networks\cite{erdHos1960} {with edge probability $p$}, (ii) Watts-Strogatz small-world (WS) networks\cite{watts1998} {with rewiring probability $q$}, and (iii) scale-free networks \cite{barabasi2003} generated by the configuration model \cite{Mo:Re95} {with power law exponent $\gamma$}.

\begin{figure}[h]
\centering
\includegraphics[scale=.8]{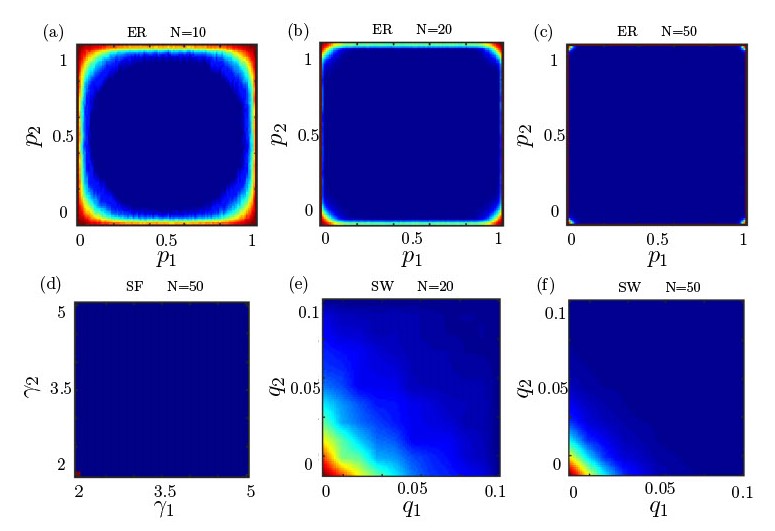}
\caption{{(a)-(c) The index $I_d$ for two Erd{\H{o}}s-R{\'e}nyi topologies {(ER)} is plotted vs the connection probabilities $p_1$ and $p_2$ with (a) $N=10$ nodes, (b) $N=20$ nodes, and (c) $N=50$ nodes. (d) The index $I_d$ for two scale-free {(SF)} networks with $N=50$ nodes is plotted vs the power-law distribution exponents $\gamma_1$ and $\gamma_2$. (e) and (f) The index $I_d$ for two Watts-Strogatz small-world {(SW)} topologies is plotted vs the rewiring probabilities $q_1$ and $q_2$ in networks with $N=20$ and $N=50$ nodes, respectively..
Different values of the index $I_d$ are shown as variation in the color spectrum from dark blue ($I_d=0$) to dark red ($I_d=1$).}}
\label{fig1}       
\end{figure}

{For each network class, we create two random graphs with the same number of nodes $N$ but with possibly two different parameters.
For each case, let $A^{(1)}$ and $A^{(2)}$ be the two adjacency matrices and $L^{(1)}$ and $L^{(2)}$ be the two Laplacian matrices.
The SBD is found using the method described above and the performance index $I_d$ is computed which is shown in Fig. \ref{fig1}.}
%
{For ER networks,} Fig.\ \ref{fig1} (a) (b) and (c) show the index $I_d$ versus  edge probabilities $p_1$ and $p_2$ for {number of nodes} $N=10, \: 20$ and $50$, respectively.
{For SW networks,} Fig.\ \ref{fig1} (e) and (f) show the index $I_d$ versus the rewiring probabilities $q_1$ and $q_2$ with $N=20$ and $N=50$ nodes, respectively. Fig.\ \ref{fig1} (d) shows the index $I_d$ versus the exponents of power-law distribution $\gamma_{1}$ and $\gamma_{2}$ of scale-free networks with $N=50$ nodes where the minimum degree of each node is set to $k=3$ in order to have a connected network.
The size of the networks in panel (a) is $N=10$ nodes, in panel (b) and (e) is $N=20$ nodes, and in panel (c), (d), and (f) is $N=50$ nodes. 
Different values of the index $I_d$ are shown as variations in the color spectrum from dark blue ($I_d=0$) to dark red ($I_d=1$). 
{For the ER networks, we see that $I_d$ is non-zero near the perimeter of the parameter space corresponding to graphs with either low edge probability or high edge probability (very sparse or very dense).
In this regime, there are many isolated nodes (sparse) or cliques (dense) which behave similarly (the graph complement of a clique is a set of isolated nodes).
These structures typically result in a finer SBD.}\\
\indent
{For the SW networks, we see only in the lower left corner that $I_d > 0$ which represents graphs that are still quite lattice-like, that is, not many edges from the original lattice have been rewired.
This means the two graphs may have large parts that are structurally identical to each other which in turn may yield more significant dimension reductions.}
Also,  Panel (d) shows that different values of the exponent of the power-law distribution $\gamma \in [2,5]$ for two scale-free networks with $N=50$ nodes results in a large dark blue area. 
{By construction, the scale-free networks we create cannot have isolated nodes (as we have set minimum degree $k = 3$) and do not have any regular structure due to the configuration model's random wiring procedure.
Thus, neither of the proposed situations which can lead to the SBD transformation significantly reducing the dimension of two random graphs (isolated nodes/cliques or shared structure) hold and $I_d = 0$ for almost all pairs of parameters $\gamma_1$ and $\gamma_2$.}
%

\section{Application of the SBD technique to cluster synchronization} \label{Ap2}
The stability of cluster synchronous solutions in networks has attracted much attention in the last few years. 
A general equation for a network of coupled dynamical systems is the following,
\begin{equation}
\dot{\bx}_{i}(t) = \bF(\bx_{i}(t))+\sum_{j=1}^{N}A_{ij}\bH(\bx_{j}(t)) \quad  i=1, \cdots, N
\label{Eq9}
\end{equation}
where 
the network connectivity is described by the adjacency matrix $A$, where $A_{ij}=A_{ji}>0$ if there is a connection between nodes $i$ and $j$ and $A_{ij}=A_{ji}=0$ otherwise. The function $\bH$ is the node-to-node coupling function. 

  The nodes of the network can be partitioned into a set of $C$ equitable clusters or balanced colors $\mathcal{C}_1,\mathcal{C}_2,...,\mathcal{C}_C$, where $N_i$ is the number of nodes in cluster  $\mathcal{C}_i$ and $\sum_{i=1}^C N_i=N$ \cite{teiler1966,schaub2016}. All the nodes in the same equitable cluster receive the same number of connections from each one of the clusters \cite{Egerstedt2012}. Among several possible equitable partitions of the network, there is one corresponding to the minimum number of clusters, which we will refer to as the minimum balanced coloring.
For any adjacency matrix $A$, the algorithm described by Belykh and Hasler \cite{Belykh2011} outputs the minimum balanced coloring very efficiently. 
Information about the minimum balanced coloring is contained in the $N \times C$ indicator matrix $O=\{O_{ij}\}$ where $O_{ij}$ is equal to 1 if node $i$ is in cluster $C_j$ and is 0 otherwise.\\

Similar to the case of complete synchronization {described previously}, given an equitable partition of the network nodes, we can define an invariant subspace for the set of Eqs.\ \eqref{Eq9}, which we call the cluster synchronization manifold. The dynamics on this manifold 
is the flow-invariant cluster synchronous time evolution \cite{golubitsky2005}  $\lbrace \bx^{s}_{1}(t), \bx^{s}_{2}(t), \cdots, \bx^{s}_{C}(t)\rbrace$, where $\bx^{s}_{1}(t)$ is the synchronous solution for nodes in cluster $\mathcal{C}_1$, $\bx^{s}_{2}(t)$ is the synchronous solution for nodes in cluster $\mathcal{C}_2$, and so on.\\
\indent
We can then define
the $C \times C$ quotient matrix $Q$ such that for each pair of clusters
$\mathcal{C}_u$ and $\mathcal{C}_v$,
\begin{equation}
    Q_{uv}=\sum_{j \in \mathcal{C}_{v}}A_{ij}  \quad i \in \mathcal{C}_{u} \quad u,v = 1, 2, \cdots, C
\end{equation}
All of the nodes belonging to the same cluster can synchronize on the quotient network time evolution $(\bx_{u}^{s}(t))$, 
\begin{equation}
    \dot{\bx_{u}^{s}}(t) = \bF(\bx_{u}^{s}(t)) + \sum_{v=1}^{C}Q_{uv}\bH(\bx_{v}^{s}(t)), \quad u=1,...,C.
\end{equation}
The question we are interested in is whether the cluster synchronous solution corresponding to the minimum balanced coloring is stable or unstable.

Stability of the cluster synchronous solution depends on the $mN$-dimensional equation,
\begin{equation}
\dot{\bz}(t) = \left[ \sum_{c=1}^{C}E_{c}\otimes D\bF(\bx_{c}^{s}(t))+A\sum_{c=1}^{C}E_{c}\otimes D\bH(\bx_{c}^{s}(t))\right] \bz(t)
\label{DeltaX}
\end{equation}
where the cluster indicator matrix $E_c$ is a diagonal matrix such that $(E_{c})_{ii}=1$ if node $i$ belongs to cluster $c$ and $(E_{c})_{ii}=0$ otherwise.\\
\indent
We note that by left-multiplying Eq.\ \eqref{DeltaX} by the matrix $\tilde{O} \otimes I_m$ where $\tilde{O}=(O^T O)^{-1} O^T$  we obtain the dynamics of the perturbation parallel to the synchronization manifold \cite{schaub2016,siddique2018symmetry,klickstein2019symmetry}. 

Similarly to Sec.\ IV, we would like to reduce the stability problem to a set of independent lower dimensional equations instead of dealing with the high dimensional problem, Eq.\ \eqref{DeltaX}. 
 Ref.\ \cite{Pecora2014} has proposed a dimensionality reduction approach based on group theory for the case of orbital clusters and shown that the {irreducible representation} (IRR) of the symmetry group can be used to block-diagonalize the set of Eq.\ \eqref{DeltaX}. Ref.\ \cite{Zhang2020} has applied the SBD method to characterize stability of any cluster synchronization pattern. {An important question is  whether the symmetry-independent approach of \cite{Zhang2020} may lead to a dimensionality reduction of the stability analysis in the broader
class of networks \cite{klickstein2018generating1} that have equitable clusters that are not merely the result of symmetries \cite{siddique2018symmetry}. Next we show 
that this is not the case.}

{ 
Following \cite{Zhang2020} we compute $P=\mathcal{SBD}(A, E_1, E_2,..., E_C)$. By applying $P$ to Eq.\ \eqref{DeltaX}, we obtain,
\begin{equation}
\begin{aligned}
\dot{\pmb{\eta}}(t) &= \left[(P^T\sum_{c=1}^{C}E_{c} P) \otimes D\bF(\bx^{s}_{c}(t)) \right.\\
&+ \left. (P^T\sum_{c=1}^{C}A E_{c} P) \otimes D\bH(\bx^{s}_{c}(t))\right] \pmb{\eta}(t),
\end{aligned}
\label{Eq10}
\end{equation}
 }
 where $\pmb{\eta}(t) = P^T \otimes I_{m} \bz(t)$. 
 Note that because both matrices $P^T A P$ and $P^T E_c P$ have the same block diagonal structure, so does $P^T A E_c P$, which becomes apparent by rewriting $P^T A E_c P = (P^T A  P) (P^T E_c P) $.
 Therefore, \eqref{Eq10} can be decomposed into lower dimensional equations,
 \begin{equation}
     \dot{\pmb{\eta}}_i(t)= \left[\sum_{c=1}^C (J_{i})_{c} \otimes D\bF(\bx^{s}_{c}(t)) + \sum_{c=1}^C (B_{i})_{c} \otimes D\bH(\bx^{s}_{c}(t))\right]{\pmb \eta}_i(t)
     \label{Eq11}
 \end{equation}
 where $(J_{i})_{c}$ and $(B_{i})_{c}$ are blocks of the same dimensions derived from the transformations $P^T\sum_{c=1}^{C}E_{c}P=\oplus_{j=1}^{n}(J_{j})_{c}$ and $P^T\sum_{c=1}^{C}AE_{c}P=\oplus_{j=1}^{n}(B_{j})_{c}$, respectively.


 
\subsection{Generating networks with assigned equitable partition}

In order to study the performance of the SBD reduction in the case of cluster synchronization, we need a method to generate a \textit{random} symmetric network with an assigned equitable partition. This can be done by using the algorithm described below.

First, assign the number of nodes in each of the $C$ clusters, $N_1, N_2,..., N_C$, where in order to enforce a trivial pattern of connectivity we pick $N_1, N_2,..., N_C$, so that no two such numbers are coprime, i.e. $\mbox{gcd}(N_i,N_j) > 1$, $i=1,...,C$, $j \neq i$. Second, we need to determine the relative indegree $d_{ij}$ of nodes in cluster $i$ from nodes in cluster $j$. Due to the assumption that the network is symmetric, the following condition needs to be satisfied
\begin{equation}
    N_i d_{ij}=N_j d_{ji}.
\end{equation}
One solution is $d_{ij}=N_j$ and $d_{ji}=N_i$, which corresponds to complete connectivity in which each node in cluster $i$ is coupled to all the nodes in cluster $j$ and vice versa. By the assumption that $N_i$ and $N_j$ are not coprimes, it follows that we can always choose other values of $d_{ij}$ and $d_{ji}$, namely,
\begin{equation}
    d_{ij}=\frac{N_j}{\alpha} \qquad d_{ji}=\frac{N_i}{\alpha},
\end{equation}
where $\alpha=\mbox{gcd}(N_i,N_j)>1$. Then, for each pair of clusters, we can randomly connect the nodes in cluster $i$ and cluster $j$ with $\frac{N_i N_j}{\alpha}$ bidirectional links. The intra-connectivity of each cluster is determined by first assigning the intra-degree $D_i$ of all nodes in cluster $i$ for $i = 1,2,...,N_C$. This should be chosen such that $N_i D_i$ is an even number and $D_i < N_i$.

The algorithm provided here generates a network with an assigned equitable partition as opposed to the algorithms to generate networks with assigned orbital partition presented in \cite{klickstein2018generating1,klickstein2018generating2}.


\subsection{Performance of the SBD technique applied to cluster synchronization}

For the case of cluster synchronization, we can also define a transformation corresponding to the trivial simultaneous block diagonalization (TSBD.) This corresponds to the transformation that separates the perturbation parallel to the synchronization manifold from the perturbation transverse to the synchronization manifold. By choosing $\tilde P=\oplus_{i=1}^C G_i$ where $G_i$ is an orthogonal matrix of dimension $N_i$ with one of its columns having entries that are all the same and equal to $\textbf{1}_{N_i}/\sqrt{N_i}$ 
we obtain the trivial simultaneous block diagonalization
\begin{equation}
    P^T A P= B_{\mbox{par}} \oplus B_{\mbox{orth}}, 
\end{equation}
where the block $B_{\mbox{par}}$ is $C$-dimensional and  
the block $B_{\mbox{orth}}$ is $(N-C)$-dimensional.
Hence, the largest block produced by the TSBD will have dimension $L=\max{(C,N-C)}$.
Thus for the case of cluster synchronization, we define the performance index, 
\begin{equation} \label{Index2}
  I_{cs} = \frac{L-Z}{L}  
\end{equation}
where $Z$ is the largest block dimension resulting from calculation of the FSBD for the set of matrices $\lbrace A, E_1, E_2,..., E_C \rbrace$ and where $A$ is the adjacency matrix and $E_1, E_2,..., E_C$ are the previously defined cluster indicator matrices. Again the index compares the performance of the FSBD with that of the TSBD. 
An index $I_{cs}=0$ indicates that the reduction achieved by the FSBD is the same as that of the TSBD. 
{As before for the index $I_d$, we define application of the SBD technique to be a success (a failure) for large (low) values of $0 \leq I_{cs} \leq 1$.}

\begin{figure}[H]
\centering
  \includegraphics[scale=.35]{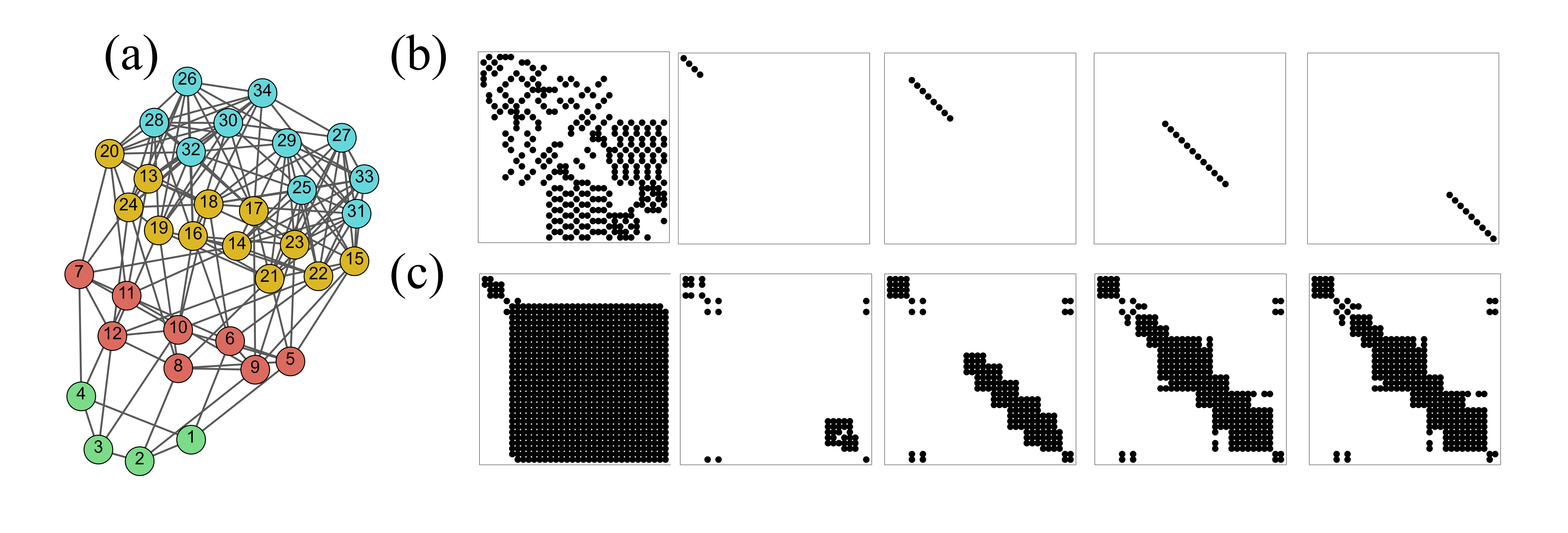}
\caption{{ (a) A randomly constructed symmetric network with $C=4$ equitable clusters and $N=34$ nodes. The clusters are identified as follows: $\mathcal{C}_1$ (green nodes) with $N_{1}=4$, $\mathcal{C}_2$ (red nodes) with $N_{2} = 8$, $\mathcal{C}_3$ (yellow nodes) with $N_{3} = 12$, and $\mathcal{C}_4$ (blue nodes) with $N_{4} = 10$. The arbitrarily chosen intra-degrees for this network are $d_1 = 2$, $d_2 = 3$, $d_3 = 4$, and $d_4 = 6$. (b) From left to right: the adjacency and cluster indicator matrices $\lbrace A, E_1,E_2,E_3, E_4 \rbrace$. Each non-zero entries of these matrix is indicated with a black dot. (c) From left to right: the block-diagonalized matrices $\lbrace P^T AP,  P^T E_1P,  P^T E_2P,  P^T E_3P,  P^T E_4P \rbrace$ after application of the FSBD transformation. }}
\label{fig2}       
\end{figure}

Next, we consider a numerical example for a random symmetric network with $C=4$ clusters generated using the algorithm described above. Figure \ref{fig2} (a) shows the network that is to be examined to measure the performance index for the SBD algorithm, with nodes color coded according to the equitable cluster to which they belong.

{
Figure \ref{fig2}(b) shows the adjacency and cluster indicator matrices for the network shown in Fig.\ \ref{fig2}(a) where each black dot represents a non-zero entry in these matrices. The block-diagonalized matrices obtained by application of the FSBD transformation are shown in Fig.\ \ref{fig2}(c). }

\begin{figure}[H]
\centering
  \includegraphics[scale=.5]{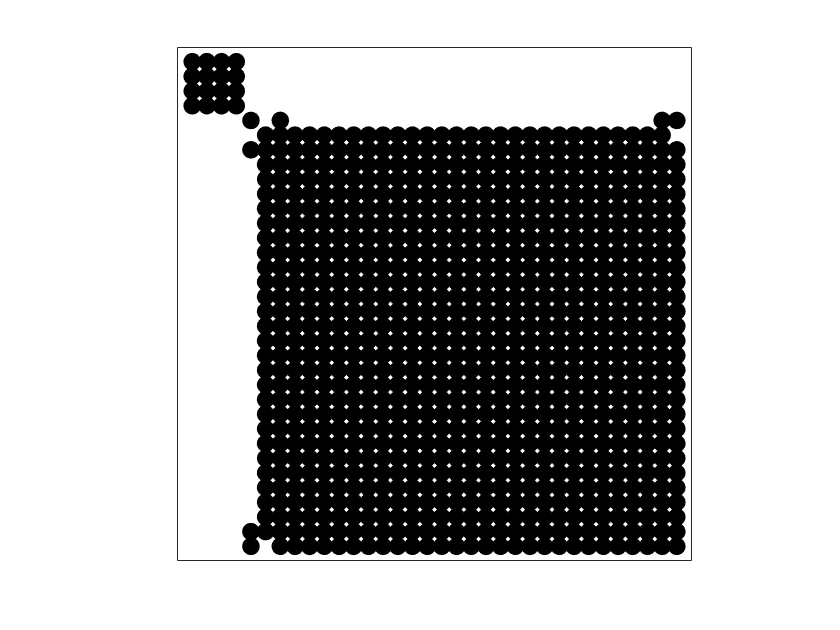}

\caption{Representation of the matrix $\Omega$ for the network shown in Fig.\ 2. The dots represent non-zero entries. Using equation \eqref{Index2}, the calculated performance index is $I_{cs}=0$ with $L = 30$ and $Z = 30$.}
\label{fig4}       
\end{figure}

In order to better visualize the block decomposition, we construct the matrix $\Omega$ as the sum of absolute values of the matrices $\lbrace P^T AP, P^T E_1 P, P^T E_2 P, P^T E_3 P, P^T E_4 P \rbrace$
\begin{equation}
\Omega = | P^T AP| + |P^T E_1 P| + |P^T E_2 P| + |P^T E_3 P| + |P^T E_4 P|
\end{equation}
where the symbol $|\cdot|$ here indicates the entry-wise absolute value of a matrix. A representation of the matrix $\Omega$ is shown in Fig.\ \ref{fig4}, which evidences {two blocks: one $4$-dimensional block and one $30$-dimensional block.}
 For this example, the calculated performance index is $0$ with $L = 30$ and $Z = 30$. We have obtained similar results for all the other instances we have tested of random graphs with assigned equitable partition (algorithm of Sec.\ VA).

\section{Discussion}\label{VI} 
{
Random `unstructured' networks have been the subject of extensive investigation in the literature, with applications to  epidemic dynamics \cite{pastor2001epidemic,marder2007dynamics,pastor2015epidemic}, percolation \cite{achlioptas2009explosive,friedman2009construction}, resilience to attacks and failures \cite{guillaume2004comparison,liu2012cascading}, games \cite{devlin2009evolution}, network synchronization \cite{restrepo2006synchronization} and control \cite{liu2011controllability}.
Several analytical results have been derived by using the assumption that the network topology is random and uncorrelated \cite{uncorrelated,restrepo2007approximating,restrepo2006synchronization,pomerance2009effect,sorrentino2019symmetries}. Complete and cluster synchronization of random networks is undoubtedly a topic of interest in the Physics and Nonlinear Dynamics literature. In this paper we take the approach of the natural scientist and focus on whether or not a mathematical tool (the SBD decomposition) is effective in dealing with the synchronization of random networks. Ref. \cite{zhang2021comment} 
takes a different perspective and claims that random networks are not a good testbed for application of the SBD technique. Here we are interested in assessing whether problems of practical interest can be successfully addressed by the SBD tool, rather than  looking for problems to which the tool can be successfully, or rather conveniently, applied. 
Previous work in this area has often only emphasized the strengths and not the limitations of the technique, which is partially corrected in this paper. The fact that the technique mostly fails when applied to random networks points out the importance of developing alternative tools and/or new techniques to deal with the important class of random networks. A relevant related question is whether the SBD technique can be successfully applied to the analysis of real network topologies. This question has been recently considered in \cite{panahi2021cluster}
, which has shown a moderate success of the SBD technique in this case.} 

\section{Conclusions}\label{s:conclusion}

The techniques for simultaneous block diagonalization of matrices have been developed by Maehara, Murota \emph{et al} in a number of seminal papers \cite{Maehara2010-1,Maehara2010-2,Murota2010,Maehara2011}. These techniques were originally applied to problems in the areas of semidefinite programming and signal processing (independent component analysis), see e.g. \cite{Maehara2010-1}. 
The first application of these techniques to network synchronization was presented in a 2012 paper \cite{Ir:So}. Only recently they have been applied to the problem of cluster synchronization of networks \cite{Zhang2020,panahi2021cluster,zhang2021unified}. 

We are highly indebted to the mathematicians who have developed the algebraic theory of simultaneous block diagonalization of matrices. This can be applied to many problems in the applied sciences where one is looking for \textit{modal decompositions} but such decompositions may not be obvious. The application of these techniques to the problem of network synchronization is important as it allows to define the extent to which the synchronization stability problem can be reduced in realistic situations that deviate from the original assumptions of nodes all of the same type and connections all of the same type \cite{Pecora1998}. We have seen here that unfortunately in generic situations (random networks) the obtained reduction is modest and comparable to that achievable with a trivial transformation. Even though that is the case, it is important to know the extent of the attainable reduction and that no further decomposition of the problem is possible. With this paper we believe we have set the expectations straight about the reduction that is realistically achievable from application of SBD to the study of complete and cluster synchronization of generic (random) graphs.  Overall, this does not diminish our enthusiasm for these techniques, which 
can provide exceptional insight into many problems of interest in physics and engineering, including network synchronization. {Besides, both Refs.\ \cite{Ir:So,Zhang2020} have shown that the reduction produced by the SBD technique can be substantial for specific network realizations, can be useful when one has the ability to appropriately select the networks connectivity. }\\

Code to compute the simultaneous block diagonalizations for the examples shown in this paper can be accessed at the Github repo \cite{Code}.

\section*{Acknowledgement}
The authors thank Prof. Kazuo Murota for insightful discussions on the subject of $*$-algebra. This research is supported by NIH grant 1R21EB028489-01A1.

 \newcommand{\noop}[1]{}

\end{document}